\documentclass[preprint2]{aastex}
\usepackage{epsfig,natbib}
\begin{document}
\setlength{\parskip}{.02in}

\title{Constraining the Adaptive Optics Point-Spread Function \\ 
in Crowded Fields: \\
Measuring Photometric Aperture 
Corrections\footnote{Based on observations obtained at the
  W.M. Keck Observatory.}}
\author{Christopher D. Sheehy, Nate McCrady\altaffilmark{\P} \\
\& \\James R. Graham {\email{jrg@berkeley.edu}} }

\affil{Astronomy Department}
\affil{601 Campbell Hall}
\affil{University of California}
\affil{Berkeley, CA 94720-3411, U.S.A.}
\altaffiltext{\P}{Now at the Department of Physics \& 
Astronomy, UCLA, Los Angeles, CA 90095}


\begin{abstract}

The point-spread function (PSF) of an adaptive optics (AO) system is
often poorly known.  This ignorance can lead to significant 
systematic errors. 
Since the degree of AO correction is sensitive to the observing
conditions: seeing, wind speed, brightness of the wavefront reference,
etc., it would be desirable to estimate the PSF from the data
themselves rather than from observations of a PSF star at another
time.  We have developed a method to estimate the PSF delivered by an
AO system in the case where the scene consists of a crowded star
field.  We model the modulation transfer function (MTF) of several key
components of the imaging system (atmosphere filtered by an AO system,
telescope pupil, and pixel array).  
The power spectrum of the image,
even a dense star field, can be used to constrain our model, which in
turn can be used to reconstruct the PSF.
In the case of circularly symmetric PSFs, we demonstrate that the
power spectrum of the source distribution function can be successfully
removed from the measured MTF and that our fit successfully recovers
input parameters from a model data-set constructed from these
parameters.  We also show that the method yields reasonable fit
parameters and a useful approximation to the PSF when applied to data
from the laser guide star (LGS) AO system at the Keck Observatory.  
Comparison of Keck LGS/AO data and Hubble Space Telescope
observations with NICMOS show that photometric accuracy of a few
percent
can be achieved for data with Strehl ratios as low as 4\%.

\end{abstract}

\keywords{atmospheric effects---
instrumentation: adaptive optics
methods: data analysis---techniques: photometric---
stars:supergiants: variable}


\section{The Adaptive Optics PSF}

Adaptive optics provides the ability to achieve near
diffraction-limited performance on large ground-based telescopes
\citep[e.g.,][]{2000SPIE.4007..115H, 2004SPIE.5490....1W, 
2000SPIE.4007...72R}.
There is a strong incentive to develop and deploy AO systems because
the cost of a large, ground-based telescope equipped with AO is modest
compared to that of a comparable aperture space telescope.  
Most large observatories already have AO systems either in place or
under development, and the data from these systems are beginning to
yield scientific results \citep[e.g.,][]{2005AJ....130.1212C,
2005ApJ...629...29M, 2005ApJ...625L..27M, 2005ApJ...621..738M,
2004Sci...303.1345P}.

While AO images show clear advantages over seeing-limited data in
resolution and sensitivity, there remain significant obstacles to the
full analysis of these data.  Characterization of the AO point-spread
function (PSF) is arguably the most serious of these challenges.
Conventional AO systems deliver modest correction (Strehl ratio, $SR
\la 0.5$) over a field of view of a few tens of arc seconds
\citep[e.g.,][]{2006ApJ...637..541F}---although some high-order
natural guide star systems afford better performance
\citep[e.g.,][]{2005AJ....130.1212C}.  Experience shows that the
chance of finding a suitable PSF star within this field is
small. Moreover, because the precision and accuracy of wavefront
correction is sensitive to the observing conditions, including seeing,
wind speed and brightness of the wavefront reference it is often a
futile exercise to observe a PSF calibration field at a different
time.

Changing atmospheric conditions and AO performance lead to variable
correction, causing variations in the amount of energy in the
diffraction-limited core of the PSF relative to the uncorrected seeing
halo.  These variations affect the Strehl ratio.  Systematic errors in
photometry will occur if these Strehl variations are not taken into
account in evaluating the photometric curve of growth and the
corresponding aperture corrections.  The angular scales of diffraction
and seeing are $\lambda/D_{tel}$ and $\lambda/r_0$, respectively,
where $\lambda$ is the observing wavelength, $D_{tel}$ is diameter of
the telescope, and $r_0$ is the Fried parameter.  For a large
telescope at a good site, operating in the near infrared (IR), the
ratio $D_{tel}/r_0 \simeq 10-20$.  Thus, the surface brightness of the
seeing halo can be more than two orders of magnitude fainter than the
PSF core, rendering it challenging to measure the seeing halo
directly---even if a suitable star is present.  The direct approach to
estimating the PSF is further frustrated because science targets
selected for AO observations will tend to have multiple sources per
seeing disk.

An innovative approach uses telemetry from the AO system to estimate
the PSF from the residual wavefront errors
\citep{veran1997PhD}. However, this method has not been implemented
for all AO system architectures, and it cannot correct 
for non-common path wavefront errors between the
wavefront sensor and the science camera. 
Myopic deconvolution is another
promising alternative, which simultaneously estimates the source
distribution and the PSF \citep{1993ApJ...415..862J}.  In the
``StarFinder'' code \citep{diolaiti}, which is designed for
analysis of variable Strehl-ratio observations of stellar fields, the
PSF is extracted directly from the image frame.

These approaches use information about the imaging system to various
degrees. For example, in myopic deconvolution it is possible to specify
the support of the PSF in frequency space
\citep{1993ApJ...415..862J}. It is our objective to capture some
rudimentary properties of the atmosphere, AO system, telescope and
detector to build a model PSF, which can be constrained by AO
observations of astronomical scenes.  Our primary interest is in
estimating the photometric curve of growth, so we consider only the
azimuthally averaged PSF. Initial astronomical application of this
algorithm was made by one of us (JRG) in
\citet{2004Icar..169..250D}.

In \S~\ref{mtfpsf} we present our description of an AO system,
including partial wavefront correction (\S~\ref{partialao}).  In
\S~\ref{extract} we describe how to interpret the power spectrum of an
AO image in terms of the MTF and the source distribution power
spectrum.  Section \ref{keckaopsf} describes the influence function of
the deformable mirror in the Keck AO system.  We use the resulting
model to simulate Keck AO data (\S~\ref{fitting}) and show that the
PSF can be recovered from synthetic images (\S~\ref{fakefitting}), and
the results of application to Keck LGS AO data are outlined in
\S~\ref{realfitting}.


\section{The Transfer Function, Point Spread Function and 
Encircled Energy}
\label{mtfpsf}

The complex optical transfer function (OTF) describes the response of
a linear optical system at some spatial frequency in the image plane,
$\nu$.  The point-spread function (PSF) forms a Fourier pair with the
system OTF.  Each independent component of an optical system, from the
atmosphere through to the detector, has its own OTF.  The system OTF
is the product of each of the separate OTFs \citep[e.g.,][]{schroeder}.  
For the bulk of this paper, we consider
only circularly symmetric PSFs.  In this case, the complex part of the
OTF is zero and the system is fully described by the real part of the
OTF, which is known as the modulation transfer function (MTF).  We
consider three principal components
that contribute to the system MTF: the
atmosphere (partially corrected by an AO system), the telescope pupil,
and the detector. Our notation is summarized in Appendix \ref{mtfs}.

\subsection{Partial Wavefront Correction}
\label{partialao}

\begin{figure}[ht!]
\begin{center}
\plotone{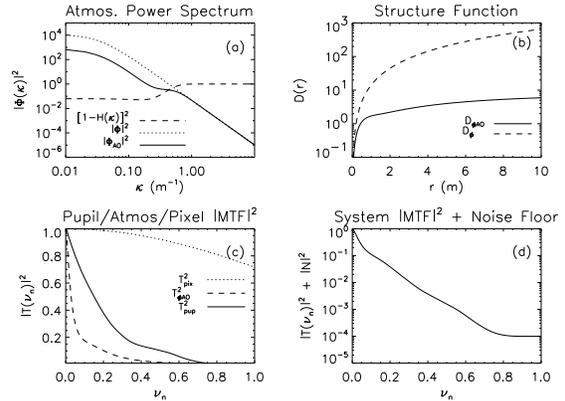}
\caption{Calculation of a model MTF for the Keck AO system with
   parameters $r_0$ = 65~cm, $L_0$ = 30.0~m, $w$ = 1.5, and
   $\sigma_{DM}$ = 56~cm: (a) Raw
  atmospheric power spectrum, $|\Phi|^2$ (dotted), the DM filter
  function, $[1-H'(\kappa)]^2$ (dashed), and the AO corrected power
  spectrum, $|\Phi_{\rm AO}|^2 = |\Phi|^2[1-H'(\kappa)]^2$ (solid).
  (b) Corresponding structure functions from panel (a) for the raw
  (dashed) and corrected (solid) atmospheric power spectra.  (c) The
  transfer functions (squared).
  $T_{\phi_{\rm AO}}$ is calculated from the AO corrected structure
  function in (b).  $T_{pup}$ is the radially averaged approximation
  of the Keck telescope's pupil MTF.  $T_{pix}$ is the pixel MTF for
  the narrow camera of NIRC2 with a platescale of $0\farcs01$
  pixel$^{-1}$.
  (d) The model system MTF,
  $T_{sys}=T_{\phi_{\rm AO}}T_{pup}T_{pix}$,  plus a constant noise
  floor. The corresponding PSF is shown in
  Figure~\ref{perf_65psf}. \label{modelmtf}}  
\end{center}
\end{figure}

Suppose the phase in the pupil plane of the telescope at a point
$\mathbf{r}$ is $\phi(\mathbf{r})$ and that the phase correction
presented by the deformable mirror (DM) at this point can be described
by the convolution
\begin{equation}
\phi_{DM}(
\mathbf{r}) 
= \int \left[ \phi (\mathbf{r}') + \delta \phi(\mathbf{r}') \right]
h(\mathbf{r}-\mathbf{r}')
d \mathbf{r}',
\label{dm_convolution}
\end{equation}
where $ \delta \phi(\mathbf{r})$ represents the error in the
wavefront, and $h(\mathbf{r})$ is the influence function of the DM,
i.e., the shape on the DM created by pushing a single actuator.  The
term $\delta \phi (\mathbf{r})$ can be used to represent a static
wavefront aberration, or the wavefront measurement error.  The
corrected phase after the DM is 
\begin{equation}
\phi_{\rm AO} (\mathbf{r}) 
= \phi(\mathbf{r}) - 
\phi_{DM}(\mathbf{r}).
\label{dm_correction}
\end{equation}
Taking the Fourier transform of Eq.~(\ref{dm_correction}),
substituting $\phi_{DM}$ from
Eq.~(\ref{dm_convolution}), and applying 
the convolution theorem, the expectation value of the power
spectrum of residual phase errors at spatial frequency in the 
pupil plane,  
$\kappa = |\mbox{\boldmath $\kappa$}|$, is
\begin{equation}
\label{phitwiddle}
|\Phi_{\rm AO}(\kappa)|^2 = 
\left[ 1-H(\kappa) \right] ^2
|\Phi(\kappa)|^2 + H^2(\kappa) \Delta_\phi^2 ,
\end{equation}
where $H(\kappa)$ is the Fourier transform of the influence function
and $\Delta_\phi^2$ is the variance of wavefront measurement error.
In deriving
Eq.~(\ref{phitwiddle}), we assumed that $\delta \phi
(\mathbf{r})$ represents uncorrelated, zero-mean noise.  The notation
$|\Phi|^2$ denotes the atmospheric phase error power spectrum (see
\S~\ref{atmomtf}).  The AO system can be thought of as high-pass
filter in the pupil plane, with filter function $\left[
  1-H(\kappa)\right]^2$ (see \S~\ref{keckaopsf} and Fig.~\ref{modelmtf} (a)).

The AO-corrected atmospheric structure function, 
$D_{\phi_{\rm AO}}(r)$,
can be found from Eq.~(\ref{DofR}) by substituting the filtered phase
error power spectrum $|\Phi_{\rm AO} (\kappa)|^2$ from
Eq.~(\ref{phitwiddle}).  Since $|\Phi_{\rm AO}(\kappa)|^2$ and
$H(\kappa)$ are both circularly symmetric the 1-d form of the integral
can be used.  The expression for the MTF of an AO-corrected atmosphere
is then given by Eq.~(\ref{DtoT}),
\begin{equation}\label{t_atmos_ao}
T_{\phi_{\rm AO}}(\nu) = \exp[\textstyle -\frac{1}{2} \displaystyle 
    D_{\phi_{\rm AO}}(\lambda f \nu)].
\end{equation}
The final system MTF is given by the product of the MTFs for the
atmosphere, the telescope (Eq. \ref{pupmtf}), and the detector
(Eq.~\ref{pixelmtfeq})
\begin{equation}
\label{sysmtf}
T_{sys}(\nu) = T_{\phi_{\rm AO}}(\nu)\; T_{pup}(\nu)\; T_{pix}(\nu).
\end{equation}
The final step involves
finding the 
PSF 
(Eq.~\ref{TtoPSF})
or the encircled
energy (EE) 
(Eq.~\ref{TtoEE})
from the Hankel transforms of $T_{sys}(\nu_n)$.


\section{Extracting the MTF from Astronomical Data}
\label{extract}

We now describe a procedure for estimating the MTF from observations
of a crowded stellar field.  Consider an astronomical scene consisting
of a distribution of sources of amplitude and position given by the
function $s(\mathbf{x})$.  The resultant image, $i(\mathbf{x})$, 
\begin{equation}
\label{imageformation}
i(\mathbf{x}) = s(\mathbf{x}) * p(\mathbf{x}) +
                n(\mathbf{x}).
\end{equation}
is the convolution of this scene with the PSF plus additive noise
$n(\mathbf{x})$. We assume that $n(\mathbf{x})$ is zero-mean Gaussian
noise.  This assumption will require our data are accurately
sky-subtracted before analysis.  Taking the Fourier transform of
Eq.~(\ref{imageformation}) and applying the convolution theorem, we
find that the power spectrum of the image, $|I(\mbox{\boldmath
$\nu$})|^2$, is the product of the source distribution power spectrum,
$|S(\mbox{\boldmath $\nu$})|^2$, and the PSF power spectrum, plus a
constant noise floor, $|N|^2$.  Since the PSF and the MTF are Fourier
pairs, the power spectrum of the image is given by
\begin{equation}\label{scene}
|I(\mbox{\boldmath $\nu$})|^2 = 
|S(\mbox{\boldmath $\nu$})|^2 \cdot |T(\mbox{\boldmath $\nu$})|^2 + |N|^2 .
\end{equation}
If the source distribution is spatially uncorrelated then 
$|S(\mbox{\boldmath $\nu$})|^2$
can be approximated as a multiplicative constant, and we should be
able to estimate the MTF directly from 
$|I(\mbox{\boldmath $\nu$})|^2$.  
Using a model for the imaging performance of the AO system
(e.g., \S~\ref{keckaopsf}), we may fit
$|I(\mbox{\boldmath $\nu$})|^2$,
yielding the MTF and
parameters that characterize the seeing and AO
performance.  With an estimate of the MTF in hand we can compute
quantities of photometric interest, e.g., the encircled energy versus
radius. 

If the distribution of sources in an image is not random then
it is not a good approximation to treat 
$|S(\mbox{\boldmath $\nu$})|^2$ as
a constant and the observed power
spectrum is not simply proportional to $|T(\mbox{\boldmath
$\nu$})|^2$.  Even if the underlying spatial source distribution is
random, astronomical luminosity functions are 
commonly steep, and
the power in an image
can be dominated by the 
the Poisson fluctuations assoicated with 
a few bright sources.  
Thus, if we do not take
account of $|S(\mbox{\boldmath $\nu$})|^2$ when we estimate
$|T(\mbox{\boldmath $\nu$})|^2$ our results will 
suffer from significant bias.

We can correct this systematic error in $|T(\mbox{\boldmath
$\nu$})|^2$ by first estimating $|S(\mbox{\boldmath $\nu$})|^2$ from
the location and {\em relative} brightness of stars in the image.  This can
be accomplished by using the methods of crowded field photometry to
build a model image, $s_{mod}(\mathbf{x})$
\citep[e.g.,][]{1987PASP...99..191S}.  The power spectrum of
$s_{mod}(\mathbf{x})$ can be used as an estimate of
$|S(\mbox{\boldmath $\nu$})|^2$, and the model MTF is multiplied by
$|S_{mod}(\nu)|^2$ prior to fitting to the observed power spectrum.
Our results 
show that even a simple estimate of $s(\mathbf{x})$
without any knowledge of
the PSF is effective.  This tactic works well,
even when the uncorrected seeing halos of multiple adjacent stars
overlap, which will be a common occurrence in AO data.  Moreover, this
procedure can be applied iteratively with DAOPHOT or StarFinder to
refine the PSF and thereby improve $s_{mod}(\mathbf{x})$.  If the
density of stars is sufficient that their diffraction-limited cores
are blended, then it is unlikely that we can measure
$s_{mod}(\mathbf{x})$ with any reliability. However, this is the
circumstance when the approximation $|S(\mbox{\boldmath $\nu$})|^2 =$
constant may be adequate. 

Phase information has been discarded in Eq.~\ref{scene} and some
information about the azimuthal structure of the PSF has been
lost. However, our goal here is to estimate the angular scales on
which is light is scattered so that we can recover the photometric
curve of growth.  Thus, for the current application we choose to
consider only the azimuthally averaged PSF and its corresponding 1-d
power spectrum.  Astronomical AO systems do exhibit non-axisymmetric
aberrations---and Eq.~(\ref{scene}) does encode information about
even-order aberrations, e.g., astigmatism. Thus, modeling the 2-d
power spectrum may be promising for describing effects such as
anisoplanatism.

The data are generally multiplied by a window function that represents
the field of view and by an apodizing function, which is necessary to
reduce spectral leakage.  In the Fourier domain, the window function
convolves the Fourier transform.  Consequently, $|I(\mbox{\boldmath
$\nu$})|^2$ is broadened relative to the PSF power spectrum by the
window function.  For a square window of width $A$, the FWHM of the
corresponding $\mbox{sinc}^2$ function is $0.8859/A$.  Since $A$ is
typically several hundred pixels, the broadening is slight and is
ignored in this treatment.


\section{Model Keck AO MTF and PSF}
\label{keckaopsf}

To explore the potential for measuring the MTF from astronomical data,
we create synthetic images similar to those obtained by an actual AO
system.  Ultimately, we will consider data obtained using the NIRC2
camera fed by the Keck Observatory's AO system operating in LGS mode.
We thus construct model AO PSFs and MTFs with parameters appropriate
for this system.

For a telescope with a circular pupil and a circular central
obscuration, $T_{pup}$ is a simple analytical function of spatial
frequency \citep[e.g.,][]{schroeder}.  Although dealing with a
circular pupil simplifies the problem, we chose to use the radially
averaged MTF derived from the approximately hexagonal Keck pupil.  We
first generated a Keck pupil image for the NIRC2 camera operating with
the {\tt LARGEHEX}
pupil stop using the IDL procedure
{\tt NIRC2PUPIL}, which is supplied by the Keck 
Observatory\footnote{\url{http://www2.keck.hawaii.edu/optics/lgsao/software}}.  
This pupil map includes the central obscuration and the spiders.
The 2-d autocorrelation function of this image is $T_{pup}$.
We then average this pupil image in equally spaced radial bins to
construct a radially symmetric pupil transfer function.  To express
$T_{pup}$ in terms of normalized spatial frequency, we define a cutoff
frequency by setting $D_{tel} = 10.99$~m, the diameter of the circle
circumscribing the Keck pupil.  The cutoff angular frequency of the
system, $D_{tel}/\lambda$, is subsequently defined for a telescope
with this pupil diameter.

To construct the AO corrected atmospheric MTF, we adopt a modified
Kolmogorov atmospheric phase power spectrum with a finite outer scale,
\begin{equation}
\label{kol_mod}
|\Phi (\kappa)|^2 = 0.0229\; r_0^{-5/3} L_0^{11/3}
\left(1+ \kappa^2 L_0^2 \right)^{-11/6}
\end{equation}
(cf. Eq.~\ref{kol_unmod})
\noindent \citep{strohbehn}. 
Measurements suggest that 
the outer scale of turbulence, $L_0\simeq$ 10--100~m
\citep{2001A&A...365..301S,
2001ApJ...554..505L}.
The form of Eq. (\ref{kol_mod})
is such that 
the presence of a finite
outer scale leaves the power unaffected, 
at $\kappa \gg 2\pi/L_0$.

We obtain the AO filtered power spectrum $|\Phi_{\phi_{\rm AO}}|^2$ by
adopting the influence function appropriate for the 349-actuator
Xinetics Inc. mirror employed in the Keck system
\citep{2004SPIE.5490....1W}.  The properties of this mirror are given
by \citet{2004AO.43..5460} who
approximate 
the
sinc-like influence function as the
difference of two Gaussians,
\begin{eqnarray}\label{keckIF}
h(\rho) & = & \frac{1}{2\pi\sigma_{DM}^2} 
              \Big{\{} \exp \left[ -(\rho/\sigma_{DM})^2/2 \right] - \nonumber \\
	& &      \frac{2}{\pi^2} \exp \left[ -2(\rho/\pi\sigma_{DM})^2 \right]
	        \Big{\}}
\end{eqnarray}
where $\rho$ is the radial coordinate on the DM,
and $\sigma_{DM}$ defines the spatial scale over
which a single DM actuator affects the wavefront. 
The Fourier transform of this circularly symmetric function is
\begin{eqnarray}
\label{FTkeckIF}
H(\kappa) & = & \exp \left[ -2(\pi\sigma_{DM}\kappa)^2 \right]   - \nonumber \\
          & &    \frac{1}{2} \exp \left[
             -(\pi^2\sigma_{DM}\kappa)^2/2\right] .
\end{eqnarray}
To mimic the effects of variable correction, we replace the influence
function by $H'(\kappa) = w H(\kappa)$.  A value of $w=0$ implies no
correction, $w = 1$ yields 6 dB, $w=1.5$ yields 12 dB, and $w=2$ gives
perfect correction at zero frequency.

Using Eq.~(\ref{phitwiddle}) for the filtered atmospheric spectrum
with an influence function described by Eq.~(\ref{FTkeckIF}), and
assuming a modified Kolmogorov spectrum Eq.~(\ref{kol_mod}), we can
calculate the atmospheric MTF by evaluating the integral
Eq.~(\ref{DofR}) for the structure function and
substituting the result in Eq.~(\ref{t_atmos_ao}). The final system
MTF is then the product of the partially corrected atmospheric MTF
with the pupil and pixel MTFs (Eq.~\ref{sysmtf}).  The corresponding
PSF for these parameters is then found using Eq.~(\ref{TtoPSF}).

As an example, we adopt values of $D_{tel}$ = 10.99~m, $r_0$ = 65~cm,
$L_0$ = 30.0~m, $\sigma_{DM}$ = 56~cm (the spacing of the DM actuators
projected onto the primary mirror), $\lambda$ = 1.65~$\mu$m, and
$w=1.5$.  Position in the image plane is converted into pixel units for
the NIRC2 narrow field camera assuming a scale of
$0\farcs01$~pixel$^{-1}$.  We assume that wavefront measurement error,
$\Delta^2_{\phi}$ in Eq.~\ref{phitwiddle}, is negligible, and set it
to zero.  The diffraction-limited PSF is computed by substituting
$T_{pup}$ for $T_{sys}$ into Eq.~(\ref{TtoPSF}).

\begin{figure}[ht!]
\begin{center}
\plotone{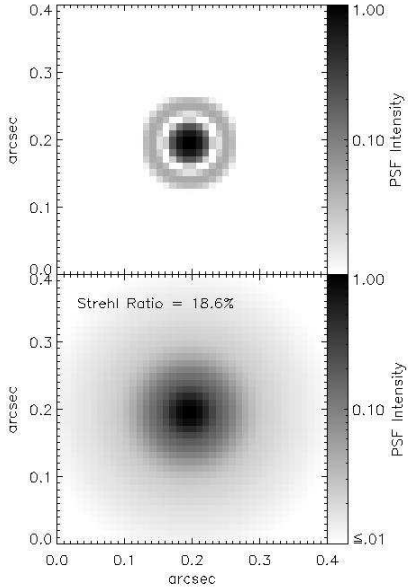}
\caption{Model 
(azimuthally averaged)
PSFs for Keck-AO/NIRC2 
narrow camera
($0\farcs01$ pixel$^{-1}$)
displayed on a
logarithmic grey scale.  Both PSFs have been scaled so that their peak
value is 1, i.e flux was not conserved. 
{\it Top:} the diffraction-limited PSF.  {\it Bottom:} the
partially-corrected PSF for $r_0=65$~cm, $L_0 = 30$~m, $\sigma_{DM} =
56$~cm, and $w=1.5$. The 
Strehl ratio is 18.6\%. \label{perf_65psf}}
\end{center}\end{figure}

Figure~\ref{modelmtf} shows the resultant filtered and unfiltered
atmospheric phase power spectra, phase structure functions, and
associated MTFs.  The filter function, $[1-H'(\kappa)]^2$ in
Eq.~(\ref{phitwiddle}), suppresses the power spectrum at low
frequencies.  Figure~\ref{modelmtf} shows that the AO system has no
effect on spatial scales smaller than the actuator spacing. The
inability of the AO system to sense and correct for phase errors above
the corresponding spatial frequency, along with its imperfect
correction within the control bandwidth, prevents the final MTF from
reaching the diffraction limited MTF.  The resultant narrow MTF
produces a PSF broadened from the diffraction-limited case with the
familiar core-halo structure in PSFs delivered by AO systems
\citep[e.g.,][]{hardy}.  The diffraction-limited PSF (no atmosphere)
and the partially corrected case are shown in Figure~\ref{perf_65psf}.
The PSFs shown are constructed by interpolating a highly oversampled
PSF onto a grid appropriate for the NIRC2 camera.  Though even a
perfect, diffraction limited PSF must be sampled onto a pixel array,
for the purposes of demonstrating change in Strehl ratio, the
diffraction limited PSF shown has been constructed without taking the
pixel MTF into account.

\section{MTF Fitting and Photometry}
\label{fitting}

\subsection{Construction and fitting of simulated data}
\label{fakefitting}

Using PSFs generated according to the recipe of \S~\ref{keckaopsf}, we
generate mock Keck images from which we then extract the MTF by
fitting our model to the resultant image power spectrum.  We evaluate
the robustness of our approach by comparing the input and recovered
parameters and we quantify photometric errors by evaluating the
discrepancy between the model and reconstructed PSF and EE.

\begin{figure}[ht!]\begin{center}
\plotone{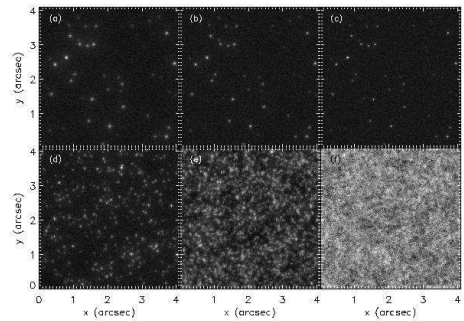}
\caption{
Model Keck-AO/NIRC2 images demonstrating the effects of
different values of $r_0$ and numbers of
stars, $N_*$. 
All images
represent $H$-band data taken with the NIRC2 camera operating at the
$0\farcs01$ pixel$^{-1}$ platescale, and share the common parameters
$L_0=30$~m, $w=1.5$, $\sigma_{DM}=56$~cm, and $\Delta_{\phi}^2 = 0$.
(a) $r_0=50$~cm, $N_*$ = 50; (b) $r_0=65$~cm, $N_*$ = 50; (c)
$r_0=1$~m, $N_*$ = 50; (d) $r_0=65$~cm, $N_*$ = 500; (e)
$r_0=65$~cm, $N_*$ = 5000; (f) $r_0=65$~cm, $N_*$ = 5$\times$10$^6$.
\label{fakeims}}
\end{center}\end{figure}

We consider three model PSFs with $r_0$ = 50, 65, and 100~cm, and
common parameters $L_0$ = 30~m, $w$ = 1.5, and $\sigma_{DM}$ = 56~cm.
Again, we ignore measurement noise by setting $\Delta^2_{\phi} = 0$.
To explore the effects of crowding, we populated images 
with
5, 50, 500, 5000, 50,000, and 5$\times$10$^6$ stars
in a field of view of $4\farcs \times 4\farcs$.
The stars have random positions and normally distributed
brightnesses. The final step of synthesizing a star field
involves convolving the array of delta functions with
the PSF. 
The
heights of the delta functions are scaled to approximate the
distribution of fluxes present in our data, and we add normally
distributed noise to the images to represent detector read noise and
Poisson sky noise so that the model images have similar
signal-to-noise ratios as our observations.  Some example model images
are shown in Figure~\ref{fakeims}.

\begin{figure}[ht!]\begin{center}
\plotone{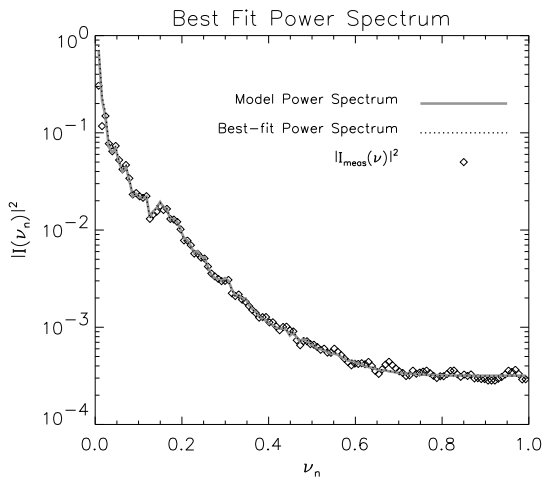}
\caption{Measured, model, and best fit power spectra for 
  the model image Figure~\ref{fakeims}(b) ($N_*$ = 50, $r_0=65$~cm).  
  The open diamonds are the power
  spectrum extracted from the image; the solid grey line is the 
  known power spectrum computed from the input data, 
  consisting of the model MTF multiplied by the
  source distribution power spectrum of the original delta function array, 
  and the
  constant noise floor; the dotted black line is the best fit power
  spectrum found after two iterations for StarFinder to estimate
  $|S_{mod}(\nu)|^2$. 
  \label{fake_data_model}}
\end{center}\end{figure}

Figure~\ref{fake_data_model} shows the measured power spectrum of the
$N_*=50$, $r_0=65$~cm image from Figure~\ref{fakeims}(b).  Also
plotted is the model MTF (squared) used to generate the PSF multiplied
by the input source distribution power spectrum, $|S(\nu)|^2$ as
calculated from the array of delta functions used to construct the
image.  The model spectrum has been fit to the data using only a
multiplicative normalization factor and an additive constant to
represent the noise.  Inspection of this plot suggests that it is
reasonable that the MTF should be recoverable from astronomical data.

Since our goal is to establish the unknown MTF for real astronomical
data we test the ability of a non-linear least squares fit to recover
the input parameters from simulated data.  We use the IDL procedure
{\tt MPFIT}\footnote{\url{http://cow.physics.wisc.edu/$\sim$craigm/idl/}} 
to perform a Levenberg-Marquardt least-squares fit of our model MTF to
Eq.~(\ref{scene}) using $|S_{mod}(\nu)|^2$ as an estimate of the
source distribution power spectrum.  Since we radially average the
measured power spectrum, we are able to directly measure the error in
each frequency bin by calculating the variance of each ensemble of
pixels. We found that reliable error estimates were necessary to
obtain unbiassed fits, especially when the AO correction was poor. 

\begin{figure}[ht!]\begin{center}
\plotone{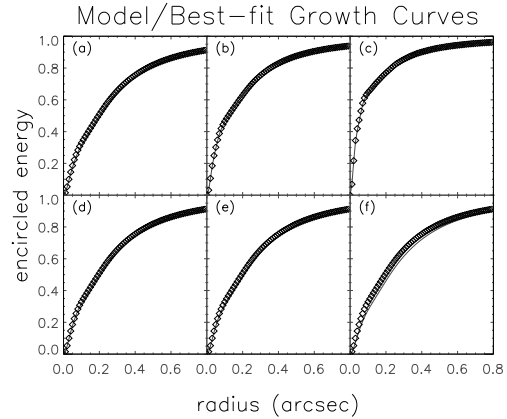}
\caption{
  Comparison of the input encircled energy curves (diamonds) with the 
  recovered encircled energy curves (solid) from the best fit MTF
  calculated from
  Eq.~(\ref{TtoEE}) for a sample of model images.  
  (a) $N_*$ = 50, $r_0=50$~cm; 
  (b) $N_*$ = 50, $r_0=65$~cm; 
  (c) $N_*$ = 50, $r_0=1$~m; 
  (d) $N_*$ = 5, $r_0=50$~cm;
  (e) $N_*$ = 500, $r_0=50$~cm;
  (f) $N_*$ = 5000, $r_0=50$~cm.
\label{meas_fitted_ee}}
\end{center}\end{figure}

The model function for $T_{sys}(\nu_n)$ in Eq.~\ref{sysmtf} is
calculated in steps. The transfer function $T_{pup}(\nu_n)$ is
calculated numerically from the Keck pupil as described in
\S~\ref{keckaopsf}. The pixel transfer function $T_{pix}(\nu)$ is
calculated analytically from Eq.~\ref{pixelmtfeq}.  Both $T_{pup}$ and
$T_{pix}$ have no free parameters.  $T_{\phi AO}(\nu)$ is, itself,
calculated in steps.  We first calculate the AO filtered atmospheric
phase error power spectrum $|\Phi_{AO}(\kappa)|^2$ as given in
Eq.~\ref{phitwiddle}, using Eq.~\ref{kol_mod} for the unfiltered
atmospheric power spectrum and Eq.~\ref{FTkeckIF} in $H'(\kappa)$.
Calculating these quantities introduces the fit parameters $r_0$,
$L_0$, $\sigma_{DM}$, $w$, and $\Delta_{\phi}$.  The structure
function is then constructed numerically using Eq.~\ref{DofR}.  (The
numerical calculation of the structure function is by far the most
computationally time consuming step, a problem compounded by the many
iterations necessary of a non-linear least squares fit.)  $T_{\phi
AO}(\nu_n)$ is then given by Eq.~\ref{DtoT}.

We consider $\sigma_{DM}$ a known quantity and hold it constant at
56~cm.  We also neglect any wavefront measurement error and set
$\Delta_{\phi} = 0$.  In addition, we fit for the multiplicative
normalization constant of the term $|S_{mod}(\nu)|^2 |T_{sys}(\nu)|^2$
and the noise in the image, $|N|^2$.  These terms, however, do not
affect the resulting PSF.

We estimate $|S_{mod}(\nu)|^2$ with a variety of methods.  The first
uses aperture photometry: for each image, we perform fits using the
{\tt DAOPHOT} routine {\tt FIND} implemented in IDL to locate
sources in the image and to estimate their relative fluxes.  The
second method is iterative: using the parameters determined from the
{\tt FIND} fits we generate a PSF and use this PSF with the crowded
field photometry package StarFinder \citep{diolaiti} to again estimate
the source function, hopefully more accurately.  (Note, we use
StarFinder only to determine the location and relative intensity of
point sources, not to estimate the PSF.)  We use parameters from this
fit to generate a new PSF to use again with StarFinder in a subsequent
iteration.  The third method is used to investigate how well the fits
work when analyzing model data and the source function is known
perfectly: we compute $|S_{mod}(\nu)|^2$ directly from the array of
delta functions that was used to construct the model image.  In either
case we create an array of size equal to the image and place delta
functions in this array at the locations of the sources with heights
equal to the fluxes.  We radially binned and averaged the resultant
2-d power spectrum to obtain $|S_{mod}(\nu)|^2$.  In order to leave
the overall measured power unchanged, we normalize $|S_{mod}(\nu)|^2$
to have a median value of 1.  $|S_{mod}(\nu)|^2$ is then an indication
of how much $|S(\nu)|^2$ deviates from being a multiplicative
constant.  Finally, we also perform fits ignoring the correction for
the source distribution power spectrum by setting
$|S_{mod}(\nu)|^2=1$.

Figure~\ref{fake_data_model} shows the best fit to the power spectrum
for image Figure~\ref{fakeims}(b) ($N_*= 50$, $r_0=65$~cm), after two
StarFinder iterations.  The final fitted MTFs are then used to compute
the PSFs and encircled energy using Eq.~\ref{TtoPSF} and~\ref{TtoEE}.
Some comparisons of the input and recovered EE curves are shown in
Figure~\ref{meas_fitted_ee}.

\subsection{Comparison of input and recovered parameters}

In general, the fits are insensitive to the model parameters for $r_0
>$ 150~cm.  For $r_0 \gtrsim$ 2~m, the MTF becomes barely
distinguishable from the perfect MTF.  In this regime it becomes
impossible to contrain any of the parameters.  It is not surprising,
therefore, that the fitted parameters for $r_0$ = 100~cm are less
accurate than those for $r_0$ = 50~cm and 65~cm.  This
degradation in accuracy applies to the fitted parameters for all of
the images, as reflected in their calculated formal errors, regardless
of the number of stars in the field.  However, in the case of high
$r_0$, the encircled energy and PSF are less dependent on the model
parameters, so the quantities of interest are left mostly unaffected.

For images containing 500 stars or fewer, the recovered photometric
growth curves deviate from the true values by less than 2\%.  The fits
for the images containing five and fifty stars generally returned the input
parameters to within the calculated error.  The recovered encircled
energy curves for these fits matched the input well, with the maximum
deviations from the known growth curves less than 1.5\%.  The accuracy
was much better than this for $r_0=65 \; cm$ and $1 \; m$.  Estimating
the source distribution power spectrum and accounting for it in the
model improved the fits and was robust.  Using simple DAOPHOT aperture
photometry routine for this purpose was sufficient, as the separation
between the stars was large.  Signal to noise ratio plays a
significant role in the accuracy of the fits.  The five star images
yielded less accurate fit parameters than the fifty star images.  Since
we did not hold the total flux constant when constructing images with
different numbers of stars, we created additional five star images scaled
to match the total signal in the fifty star images.  Performing the fits
on these new images yielded more accurate results, in line with those
obtained from the fifty star images.

It is more difficult to estimate the source distribution power
spectrum for the images with more than fifty stars because of increased
crowding.  Nonetheless, fitting the 500 star images returned
parameters that were close to the input values when we used the
iterative method to refine the estimate of the source distribution
function.  The resulting photometric growth curves, though less
accurate than those for five and fifty stars, do not deviate from the known
ones by more than 2\%.  Using the known source distribution function
for these synthentic data yields improved fit parameters, confirming
that with this many stars uncertainty in $|S_{mod}(\nu)|^2$ is the
dominant source of error.

For the images containing $\ge$ 5000 stars, the accuracy of the returned
parameters drops again.  Nonetheless, the best-fit photometric curves
of growth differ from the true ones by usually less than 5\%.  When a
field contains so many stars that every beam contains multiple stars,
estimating the spatial distribution power spectrum becomes impossible.
In this case, we hope that the assumption of a random distribution of
point sources leading to a simplification of $|S(\nu)|^2$ as a
constant becomes increasingly valid.  For the images containing 5000
stars ($\simeq$ 0.4 stars per $(\lambda/D)^2$ beam), the best fit
parameters deviate significantly from the input parameters.  The field
is sufficiently crowded that errors in $|S_{mod}(\nu)|^2$ are large,
but not so crowded that $|S(\nu)|^2$ may be approximated as a
constant.  The deviations in the fitted encircled energy from the true
encircled energy are as great as 5\%.  These same problems apply to
the images with 50,000 stars ($\simeq$ 4 stars per beam), but to a
greater extent.  In this regime, the fits almost completely fail to
return the input parameters.  Accurately estimating the source
distribution here is almost impossible.  The fits performed while
considering $|S(\nu)|^2$ as a constant do not produce better results,
indicating that the field is not random enough to allow this
assumption.

The 5 $\times$ 10$^6$ image ($\simeq$ 400 stars per beam) defies any
attempt to estimate the source distribution function.  However, in
this limiting case, the field is so crowded that neglecting to account
for $|S_{mod}(\nu)|$ does not cause the fits to fail.  The fits,
though not as accurate as those for the images with 5--500 stars,
yield encircled energies no more than 6\% different from the true
curves.

Though poor sky subtraction degrades the accuracy of the fits, the
effect can be mitigated by filtering out the low spatial frequency
bins.  We estimated the background directly from the model images and
sky subtracted before finding the power spectrum.  Since the sky
subtraction will never be perfect and is, for the images with more
than 500 stars, significantly inaccurate, the power spectrum is
contaminated.  However, if we assume that the background does not
contain much power beyond the lowest few frequency bins, we can
exclude these bins from our fits.  (The first two frequency bins in
Figure~\ref{fake_data_model} demonstrate the effect of imperfect sky
subtraction.)  For the 5, 50, and 500 star images we ignored the first
two frequency bins ($\nu_n < 0.02$), which yielded more accurate
results than using all data points.  For the 5000 star image, we
repeated the fits, filtering out the first three bins.  This resulted
in best-fit parameters somewhat closer to the inputs, but did not
significantly improve the accuracy of the best fit EE curve. 

In general, it appears that the benefits of increased signal to nosie
in the images with large numbers of stars, as well as the benefits of
the increased validity of the assumption of randomness in the source
function, are outweighed by the increased ability to sky subtract and
to estimate the source function in images with less crowding.
(Assuming, of course, that there is enough signal to be able
to use fewer stars.) 
For instance, the measured power spectra for the $N_*=5 \times 10^6$
images are noisier than for those with 5, 50, and 500
stars, and the fitted parameters are consequently 
less accurate, even though the assumption of a random spatial
distribution holds better for these images than for those with less
stars and the signal to noise is much higher. 

While our most crowded images ($N_*=5 \times 10^6$) yielded fits less
accurate than our least crowded images ($N_* \leq 500$), they still
led to best-fit curves of growth that were accurate to $<$ 6\%.  There
appears to be a regime, however, in which the field is not crowded
enough to allow for the approximation of $|S(\nu)|^2$ as a constant,
but too crowded to allow for an accurate estimation of it.  Our
example of this is the failed fits for images containing 50,000
stars.  In this case, we were unable to find a satisfactory way to
improve the fits.


\subsection{Fitting laser guide star Keck AO data}
\label{realfitting}

The nearby dwarf starburst galaxy NGC 1569 ($v_{helio}$ =
$-$104~km~s$^{-1}$) was observed with the Keck LGS/AO and the NIRC2
camera on 2005 March 02.  The field is centered on the superstar
cluster SSC A1/A2 \citep{2001AJ....122..815O}.  The NIRC2 pupil stop
was in the {\tt LARGEHEX} position (circumscribed circle) and the
camera was operating in the narrow field mode ($0\farcs 01$ per
pixel).  Observing conditions were photometric, and the seeing was
$0\farcs4$ at $H$.  The $R \simeq 11$~mag. equivalent laser guide star
was projected at the center of the science camera field of view and
SSC A1/A2 was used as the tip-tilt reference.  The science field was
observed at an airmass of 1.5 with the AO system running at 450 Hz.
We use a 60-second $H$-band exposure from this observation to
illustrate the application of our PSF estimation method.  Additional
observations with 120-s exposures were also obtained yielding a 
mosaic with a total exposure time of 15 minutes.

\begin{figure}[ht!]\begin{center}
\plotone{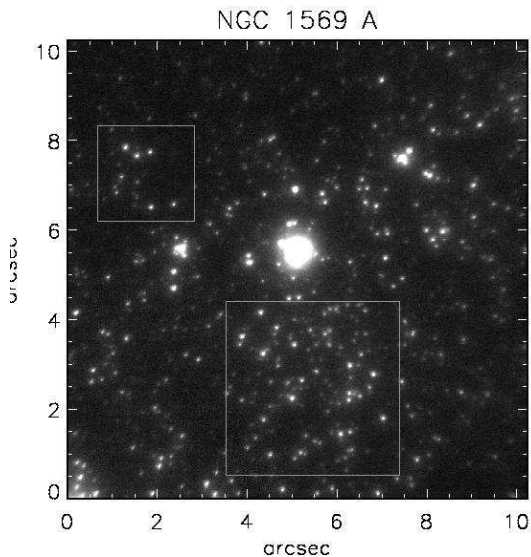}
\caption{A 60 second Keck LGS AO NIRC2 $H$-band image of
  super-star cluster A in NGC 1569.  An MTF was measured and
  fitted for the entire image and for the two indicated subarrays.
  \label{realimage}} 
\end{center}\end{figure}

Flat-fields were taken at the zenith near midnight and median filtered
to remove starlight.  Bad pixels in the single 60-s image were
repaired by substituting them with an interpolation of the surrounding
pixels.  As the crowding in the field is relatively low and displays
little variation across the field, conventional sky subtraction was
applied using the IDL implementation of the DAOPHOT routine {\tt MMM}.

\begin{figure}[ht!]\begin{center}
\plotone{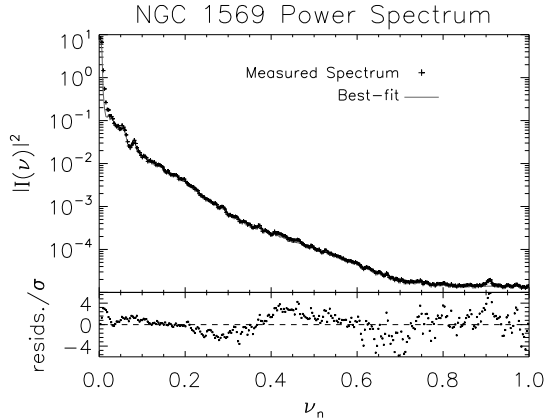}
\caption{
The measured power spectrum of the entire image
  in Fig.~\ref{realimage} and the best-fit model power spectrum
  based on the MTF and source distribution function. $\sigma$ is the
  rms error of the ensemble of pixels that was radially binned and
  averaged to produce the 1-d spectrum. 
  The fit parameters are listed in the last column of
  Table~\ref{bestparams_real}. The corresponding
  encircled energy curve is shown in Fig. \ref{realmtf_b}.  
\label{realmtf_a}}
\end{center}\end{figure}

Figure~\ref{realimage} shows the entire NIRC2 field of view.  The
image is dominated by the resolved cluster but also includes several
hundred unresolved stellar sources.  The cluster light includes
extended nebular emission and is therefore not suitable for measuring
the MTF. Our goal is to estimate the PSF for the regions containing
stars.  Therefore, we define two subarrays within Fig.~\ref{realimage}
that avoid SSC A and contain approximately 190 and 20 point sources,
respectively.  (Preliminary photometry indicates that the point
sources in the image have absolute $H$-band magnitudes consistent with
single red supergiant stars.)  Anisoplanatism is evident in
Fig.~\ref{realimage}.  We have chosen regions small enough ($\sim$
$2'' \times 2''$ and $\sim$ $4'' \times 4''$ ) so that this effect
should be negligible.  For comparison, we also compute the power
spectrum for the entire image after masking out the central cluster
and two other bright, unresolved objects.

\begin{figure}[ht!]
\begin{center}
\plotone{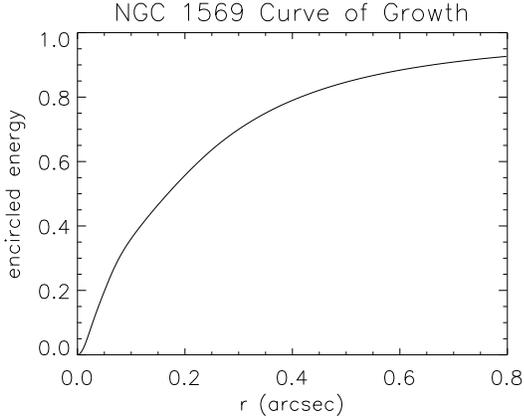}
\caption{
The recovered 
encircled energy curve
corresponding to the 
fitted power spectrum 
(see Fig.~\ref{realmtf_a})
of the entire image
in Fig.~\ref{realimage}.
The fit parameters are listed in the last column of
Table~\ref{bestparams_real}.  
\label{realmtf_b}}
\end{center}
\end{figure}

The best-fit power spectrum and encircled energy for the entire image
are shown in Figures~\ref{realmtf_a} and 
\ref{realmtf_b}.  
The best-fit parameters are listed
in Table~\ref{bestparams_real} along with those for the fits using the
two subarrays.  We again hold $\sigma_{DM}$ at a constant 56~cm.  All
parameters represent the best-fit after two iterations of using
StarFinder to estimate the source distribution power spectrum.  We
also state the formal errors in the parameters as computed from the
covariance matrix.  To account for any imperfect sky subtraction, we
do not fit data at $\nu_n < 0.02$.  Thus we filter our spectra below
the first 1, 2, and 5 frequency bins in the small and large subarrays,
and in the entire image, respectively.

\begin{figure}[ht!]\begin{center}
\plotone{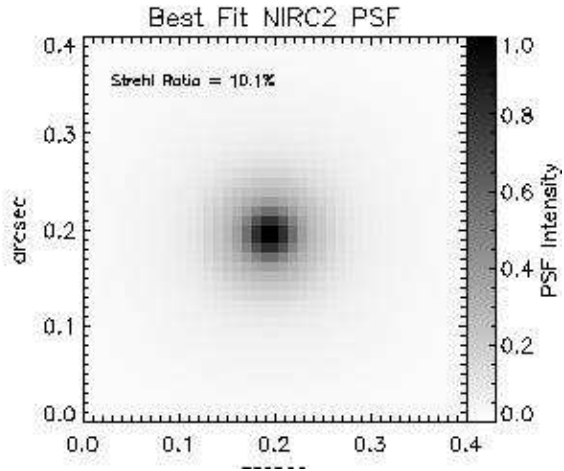}
\caption{The PSF generated 
for the entire image's best-fit parameters
listed in Table~\ref{bestparams_real}. 
\label{fittedpsf}}
\end{center}\end{figure}

In \S~\ref{fakefitting} our model PSFs were circularly symmetric and
based solely on our model MTF.  All we verified there was the ability
to successfully {\em extract} the MTF from data.  Since real PSFs will
contain asymmetries and since our model will not describe {\em all} of
the factors contributing to an actual MTF (i.e., deviations from
Kolmogorov turbulence, directional dependence in $D(r)$, systematic
errors in the wavefront sensor) we cannot expect our real fits to be
as accurate as our fake fits.  Nevertheless, Figure~\ref{realmtf_a}
shows excellent agreement between measured and best-fit model power
spectra---we have evidently included enough detail in our model to
capture the essence of the Keck AO system.  Based on the analysis of
our synthetic data, the recovered values put us in the regime where
the fits show good sensitivty to $r_0$ and $w$.  Additionally, the
three sets of fit parameters mostly agree to within the formal errors.
The values of $L_0$ show some discrepancy, but in this region of
model parameter space the MTF and EE show no signficant variation when $L_0$
changes by as much as 10\%.  The recovered EE curve is most sensitive
to changes in $r_0$, which stays within error over the three fits.
The density of stars in our
image is low enough to enable accurate sky subtraction and estimation
of the source distribution function.  All of these factors give us
confidence in our derived parameters.

\begin{figure}[ht!]\begin{center}
\plotone{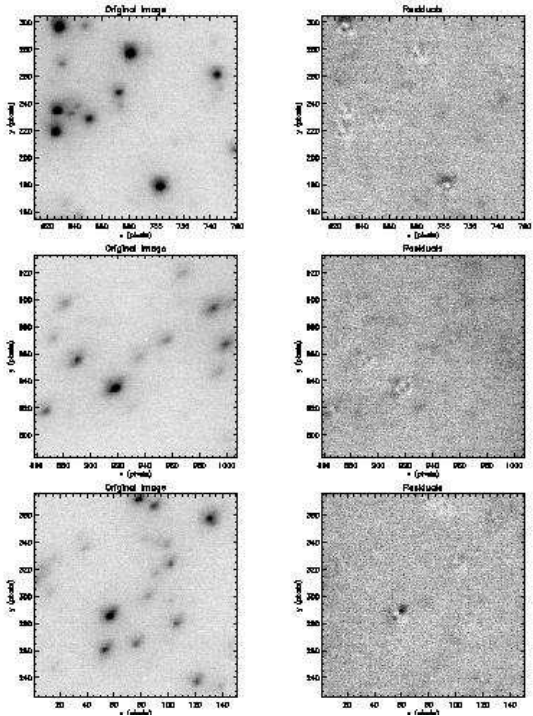}
\caption{Typical residuals after performing PSF subtraction on stars
  in Figure~\ref{realimage} using the PSF in Figure~\ref{fittedpsf}.
  All images are displayed on a linear scale and have a $1\farcs50
  \times 1\farcs50$ field of view.  The residuals show
  mainly azithumal structure, an effect we may attribute mainly to the
  failings of our model to describe non-radially symmetric PSFs.  Such
  residuals should have little effect on the encircled energy. 
  \label{resids}} 
\end{center}
\end{figure}

The formal errors decrease significantly as the size of the
subarray we use to measure a power spectrum grows.  For one, the
signal to noise ratio of the power spectrum increases when there are
more stars.  Also, larger images give finer sampling of the power
spectrum.  For these reasons, and since the best-fit power spectrum for
the whole 60-s image is a good quantitative match with the measured one and
does not differ significantly from either two subarrays, we choose the
entire image best-fit MTF to compute a PSF and EE curve.

We use the best-fit parameters to generate an MTF and compute the
corresponding PSF using Eq.~(\ref{TtoPSF}).  The recovered PSF is
shown in Figure~\ref{fittedpsf}.  This PSF exhibits the same core-halo
structure that is evident in the data.  We performed PSF fitting and
subtraction for a number of bright sources in the image.  A few of the
original stars and the residuals after PSF subtraction are shown in
Figure~\ref{resids}.  The residuals contain approximately $\pm 10\%$
of the original starlight, a failing that can probably be attributed
to our initial assumptions about the symmetry of the PSF.  Indeed,
most of the residuals exhibit asymmetric structure.  This indicates
that the actual PSF's deviation from circular symmetry, something for
which our model is unable to account, is responsible for much of the
residual starlight.

\begin{table}[ht!]
\begin{center}
\begin{tabular}{cccc} \hline\hline
         & \emph{$N_*$ $\simeq$ 20} & \emph{$N_* \simeq$
    190} & Full Image\\  \hline
\emph{$r_0\;(m)$}  & 0.58 $\pm$ 0.09  & 0.56 $\pm$ 0.04 & 0.56 $\pm$ 0.02  \\ 
\emph{$L_0\;(m)$}  & 17.5 $\pm$ 1.0   & 20.0 $\pm$ 0.6  & 18.1 $\pm$ 0.2   \\ 
\emph{$w$}         & 1.33 $\pm$ 0.08  & 1.34 $\pm$ 0.04 & 1.33 $\pm$ 0.02
    \\  
\hline
\end{tabular}
\end{center}
\caption{Best fit parameters for the two sub-arrays indicated in
  Figure~\ref{realimage} and for the entire
  image. \label{bestparams_real}} 
\end{table}


\subsection{Photometry}
\label{photometry}

In this section we investigate the accuracy of photometry extracted
from LGS/AO observations by comparing the results with data obtained
using Hubble Space Telescope/NICMOS.  The PSF for HST instruments
is well characterized \citep[e.g.,][]{1998PASP..110.1046K} hence
the aperture corrections are known and HST data provide a good
standard against which ground-based AO measurements can be compared.

Ideally, we would perform the LGS/AO photometry in two steps. First,
perform relative photometry using PSF fitting with an empirical 2-d
PSF, which is built from stars judged to be isolated.  Second, use our
1-d PSF to compute the aperture corrections and place the relative
photometry on an absolute scale. Since we are interested in the
accuracy attainable with our method we have chosen to perform PSF
fitting photometry with the model 1-d PSF obtained as described in
\S~\ref{realfitting}. While crowding will limit the precision of this
approach, there is no ambiguity in how to calculate the aperture
correction.  When applied to an ensemble of stars this one-step method
should give a reliable way of evaluating the accuracy of the resultant
photometry.

\begin{figure}[ht!]\begin{center}
\plottwo{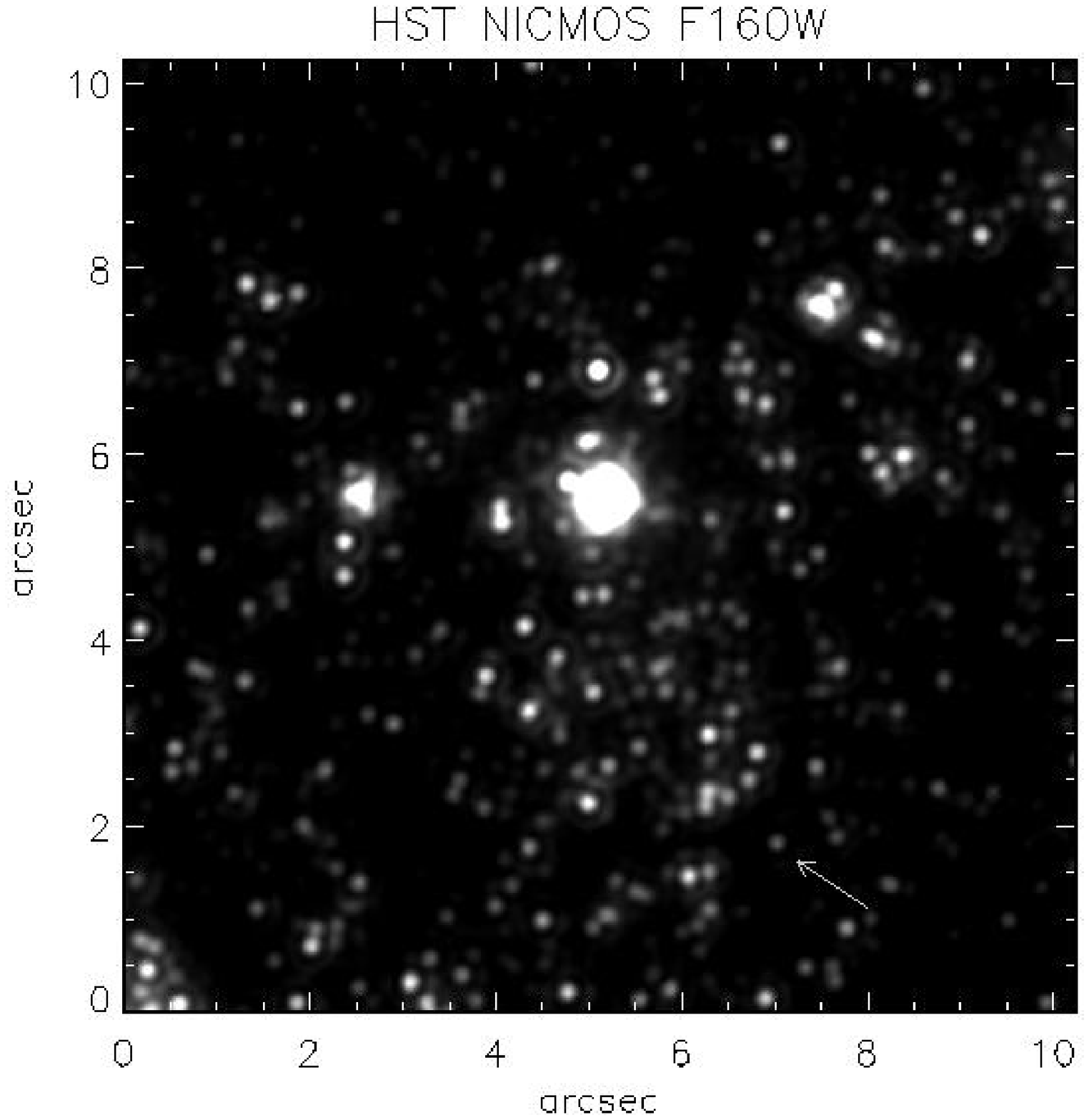}{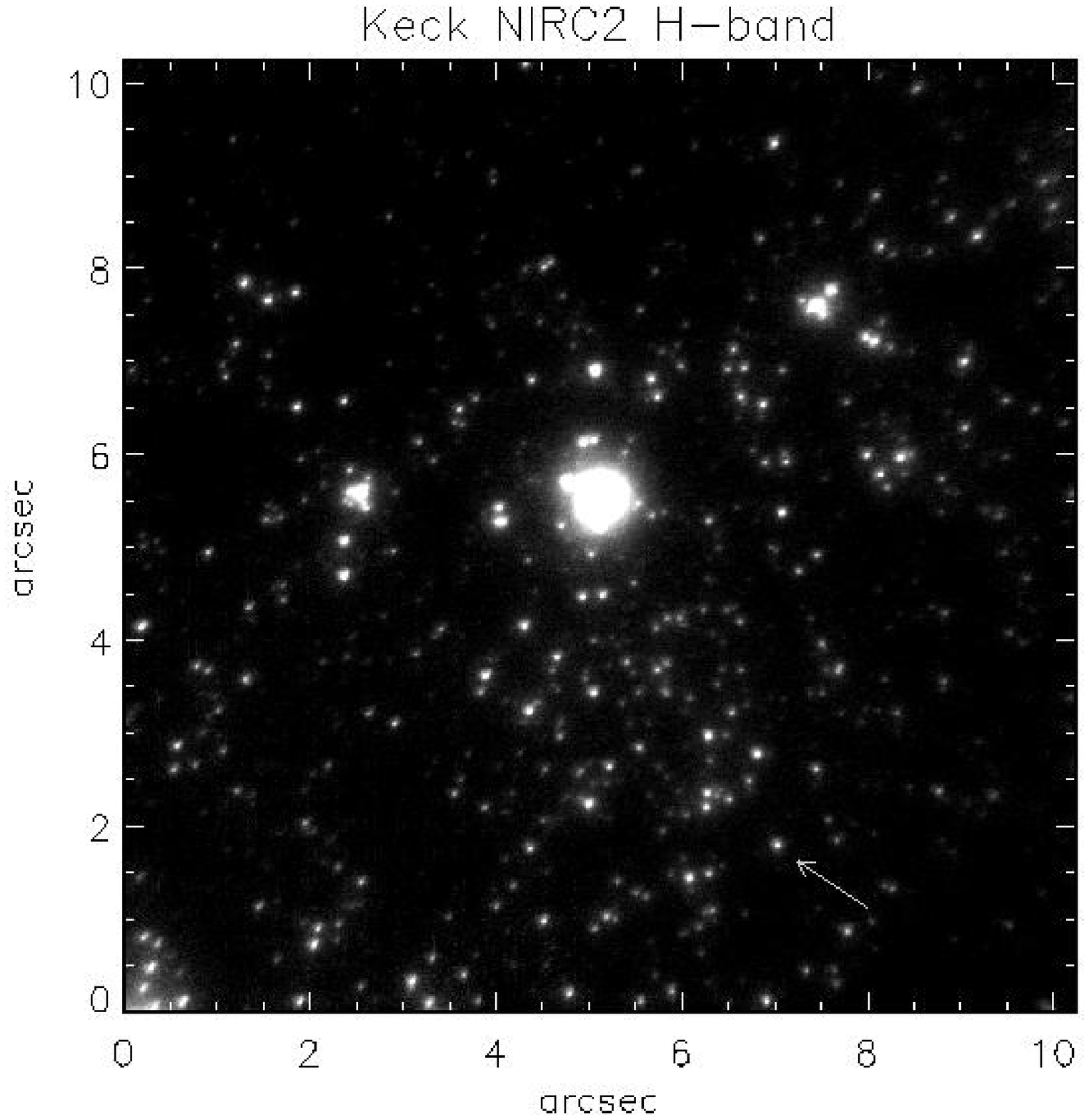}
\caption{The Keck AO image of NGC1569A and the corresponding sub-image
  of the NICMOS mosaic.  The NICMOS image has been rotated 
  to match the Keck field-of-view. The arrow indicates 
  a variable star with the largest observed fluctuation ($\delta m$ = 0.8 mag.).
  \label{keck_hst_ims}}
\end{center}\end{figure}

We measure $H$-band magnitudes for approximately 250 bright point
sources in the field of a single 60-s exposure of NGC 1569 SSC A
($r_0$ = 56~cm; Strehl = 10.1\%; see \S~\ref{realfitting}).  We
used five images of the near-infrared standard star SJ~9139
\citep{1998AJ....116.2475P} obtained on the same night as the science
data to measure the photometric zero point.

\begin{figure}[ht!]\begin{center}
\plotone{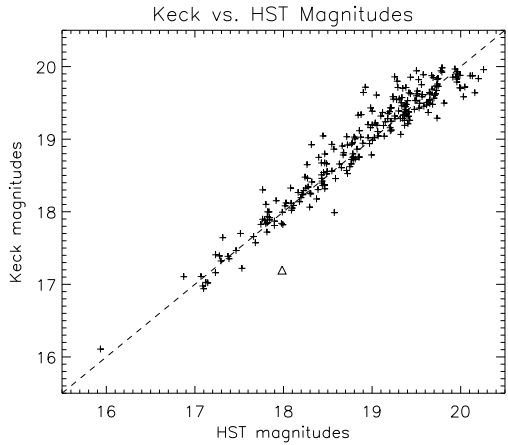}
\caption{The magnitudes for 243 stars common to the Keck and HST
  fields of view.  The HST magnitudes were taken from
  \cite{2001AJ....121.1425A} and color corrected from the NICMOS F160W
  filter to Keck's $H$ filter.  The dashed line has slope 1 and
  intercept 0.  The triangle is the star indicated by the arrows in
  Figure~\ref{keck_hst_ims}. 
  \label{keck_hst_photom}}
\end{center}\end{figure}

HST data were obtained with the NIC2 camera in the F160W filter on
1998 February 25.  The Keck image and the matching NICMOS sub-image
are shown side-by-side in Figure~\ref{keck_hst_ims}.  Photometry for
the HST/NICMOS images has been extracted by
\citet{2001AJ....121.1425A}.  Comparison of the photometry obtained
from our LGS/AO data begins with matching the catalogs of Keck and
HST/NICMOS sources.  Inspection of Figure~\ref{keck_hst_ims} shows
that Keck frequently resolves stars that appear as single sources to
HST/NICMOS.  Although we are able to identify more stars in the Keck
image than presented here, we have selected only those stars that are
separated from neighbors by more than $0\farcs 1$ in both catalogs.
The central wavelengths of the NICMOS F160W filter (1.60~$\mu$m) and
the NIRC2 $H$-band filter (1.63~$\mu$m) are similar. However, the
filter bandpasses differ significantly, and for these red stars a
color correction is necessary to compare space and ground-based
measurements.  We correct the HST F160W magnitudes to the Keck system
based on their F160W-F110W colors, using the method outlined in the
appendix of \cite{2000AJ....120.1779M}.

A plot of Keck vs. color-corrected HST magnitudes for isolated stars
(Figure~\ref{keck_hst_photom}) shows that the mean photometric
discrepancy between the two data sets amounts to only a few percent.
In this figure the data points fall along a straight line of unit
slope and intercept zero.  The mean Keck magnitude minus the HST
magnitude, $\delta m$, is $\overline{ \delta m}$ = 0.05~mag., and
standard deviation is $\sigma_{\delta m}$ = 0.21~mag.  The standard
error of $\delta m$ is 0.013 mag.

\begin{figure}[ht!]\begin{center}
\plotone{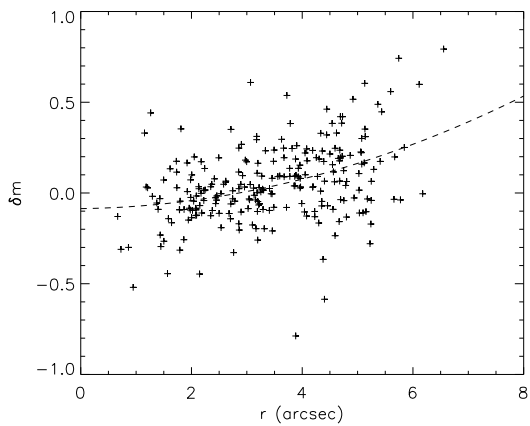}
\caption{
  The difference between the measured 60-s Keck exposure magnitudes and
  the HST magnitudes as a function of distance from the center of the
  FOV. Anisoplanatism would appear as an error that increases with 
  distance from the laser beacon (projected at $r = 0$). 
  The dashed line shows the best-fit second order polynomial, illustrating 
  that there is only a weak trend in the photometric errors. 
  \label{anisotropy}}
\end{center}\end{figure}

The agreement between the Keck and HST photometry at the 5\% level is
encouraging, and provides significant validation for our method of
finding AO aperture corrections.  However, the roughly 20\% scatter
for individual measurements, which is several times larger than that
expected from source, sky and detector noise, merits some additional
comment.

One possible source of scatter is anisoplanatism in the AO images. A
systematic change in the PSF and across the field of view would result
in seemingly random errors distributed among stars of all
magnitudes in Figure~\ref{keck_hst_photom}.  However, a plot of
$\delta m$ vs. field angle (Figure~\ref{anisotropy}) shows only a weak
trend, and indicates that variation of the AO PSF across the Keck
field of view only contributes a fraction
($\simeq$ 10\%) of the 0.21 mag. scatter.

We investigate our internal errors by comparing two independent sets
of LGS/AO observations of the field of NCG 1569 SSCA.  For the first
image we use the same 60-s exposure as above and a 120-s exposure for
the second ($r_0$ = 42~cm; Strehl = 3.7\%.  
Each image has a different PSF and a sky background
level. Comparing the photometry done from the 60s and 120s exposures
gives $\overline{m_{60s}-m_{120s}}=0.04$~mag.  and
$\sigma_{(m_{60s}-m_{120s})}=0.10$~mag.  Thus, our measurement errors
account for only about 20\% of the variance between the Keck and HST.

Likely, the major source of the observed scatter is the natural
variability of red supergiants (RSGs), the main stellar component
present in this field.  Visual inspection of the NIRC2 and NICMOS
images reveal several stars that have brightened or dimmed betweeen
1998 and 2005.  The most variable star, with $\delta m = 0.8$, is
indicated by the triangle in Figure~\ref{keck_hst_photom} and the
arrow in Figure~\ref{keck_hst_ims}.  The magnitudes and F160W-F110W
colors ($\approx 1$) of our sample indicate they are mostly RSGs with
masses $\gtrsim 10M_{\odot}$ \citep{2001AJ....121.1425A}.  The most
massive few stars are probably between $20M_{\odot}$ and
$30M_{\odot}$.  The stars in our sample are unstable and can fluctuate
significantly in luminosity.  \cite{Maeder} has studied this
variability in a sample of 2420 RSGs.  The results indicate that stars
of these masses and spectral types have average peak-to-trough
$V$-band amplitudes between 0.1 for $10 M_{\odot}$ stars and 0.4-0.8
for $30M_{\odot}$ stars.  Although the amplitude of variation is
likely to be less at infrared wavelengths, these findings suggest that
we can attribute most of the observed variance to the stars
themselves.

Since the 120-s exposure exhibited poorer image quality, it affords the
opportunity to examine the photometric integrity for reduced AO
correction.  The same photometric Keck-HST comparison for these data
give $\overline{\delta m}=-0.03$ mag. and $\sigma_{\delta
m}=0.23$~mag., confirming that accurate results
can be extracted
at low Strehl ratios.


\section{Discussion}

We have developed a simple model to describe the imaging performance
of an AO system. The model makes assumptions about the shape of the
power spectrum of atmospheric phase errors, the pupil geometry, and
the DM influence function and has a few adjustable parameters,
including the Fried parameter ($r_0$), the degree of AO correction
($w$), and the wavefront measurement noise ($\Delta\phi$).  Beyond the
inherent value of a well defined framework to quantify an observed
phenomenon, this method has several advantages in the practical
analysis of AO data.  Generating numerical MTFs is a cheap and easy
way to predict system performance under a variety of conditions, i.e.,
seeing, pupil obscuration, DM actuator spacing, \&c.

By considering the Fourier information present in astronomical images
we have shown that it is practical to fit model MTFs to image power
spectra, constrain the model parameters and recover the PSF,
encircled energy and Strehl ratio.  High fidelity simulations of AO
performance can be performed using computationally intensive Monte
Carlo codes \citep[e.g.,][]{jolissant}.  However, we believe that we
have shown that it is unnecessary to employ these elaborate methods to
compare with observations to estimate photometric curve of growth
corrections. 

Using the method outlined here, the azimuthally averaged PSF and hence
the photometric curve of growth can be estimated directly from an
image containing multiple point sources.  Data obtained in varying
observing conditions and with changing AO correction can be analyzed so
that they are not subject to systematic photometric errors, and separate
and potentially unreliable PSF star observations are not required.
Our method allows one to estimate the PSF, even if there are no
isolated stars in the image.  Although only the azimuthally averaged
PSF is derived, the fact that the residuals in Figure~\ref{resids}
show mainly azimuthal components means that the curve of growth is
valid.  A comparison of Keck LGS/AO and HST/NICMOS observations of the
same field at 1.6~$\mu$m in NCG~1569 shows that our photometric method
is accurate to 5\% or better.  The precision of our 1-d PSF fitting
photometry (about 10\%) is not limited by photon statistics and
detector noise. We are investigating a hybrid approach which involves
conventional 2-d PSF fitting photometry using an empirical kernel
followed by application of aperture corrections derived from MTF
fitting.

Treating the OTF as a one-dimensional MTF is an approximation. The
wavefront delivered by an AO system often suffers from systematic
calibration errors which lead to significant PSF azimuthal
asymmetry. Non-circular pupils and anisotropic atmospheric turbulence,
e.g., due to a prevailing wind, are the rule rather than the
exception.  While odd-order aberrations, e.g., coma, cannot be
reconstructed from the power spectrum, even aberrations like
astigmatism can be described.  A natural extension to the work
presented here is to fit a 2-d OTF to the 2-d power spectrum to capture
this information.  It will be particularly interesting to investigate
the ability of this method to describe the astigmatism introduced by
angular anisoplanatism by mapping out the variation of model parameters
as a function of field angle. 

\begin{acknowledgements}

We thank Melissa Enoch for her contributions to developing an early
version of this algorithm that served as a reference point from which
to explore the topic further.  We are also grateful to Alessandra
Aloisi for her assistance in verifying the accuracy of our method by
making available to us the HST photometry of NGC~1569.

This work has been supported by the National Science Foundation
Science through AST 0205999 and the Center for Adaptive Optics,
managed by the University of California at Santa Cruz under
cooperative agreement No. AST 9876783.  The W.M. Keck Observatory is
operated as a scientific partnership among the California Institute of
Technology, the University of California, and the National Aeronautics
and Space Administration.  The Observatory was made possible by the
generous financial support of the W.M. Keck Foundation.  In addition,
the authors wish to recognize and acknowledge the very significant
cultural role and reverence that the summit of Mauna Kea has always
had within the indigenous Hawaiian community.  We are most fortunate
to have the opportunity to conduct observations from this mountain.

\end{acknowledgements}

\appendix
\section{Appendix: Modulation Transfer Functions}
\label{mtfs}

The optical transfer function (OTF) specifies the amplitude and phase
response of a linear optical imaging system to a stimulus at a given
spatial frequency.  In general the OTF is complex. Here we consider
the special case of a circularly symmetric PSF, in which case the
complex OTF reduces to the real modulation transfer function (MTF).
Following \citet{schroeder}, the MTF, $T(\nu)$, is a measure of the
contrast at spatial frequency $\nu$ in the image, relative to the
object.
We expect $T(\nu)\rightarrow 1$ as $\nu \rightarrow 0$. For a pupil
with finite support, above some critical spatial frequency, $\nu_c$,
$T(\nu_c)= 0$ and all information at frequencies greater than $\nu_c$
is lost.

In the image plane, the cutoff spatial frequency of a telescope with a
circular pupil, diameter, $D_{tel}$, and observing wavelength,
$\lambda$, is
\begin{equation}\label{nuc}
\nu_c = 1/F\lambda.
\end{equation}
This corresponds to a linear image size of $f\lambda/D_{tel} =
F\lambda$, where $f$ is the effective focal length at the image plane
and $F=f/D_{tel}$ is the focal ratio. It is convenient to use spatial
frequencies in units of the cutoff frequency in the image plane, or
normalized spatial frequencies, $\nu_n = \nu / \nu_c = \nu \lambda F$.

\subsection{Atmospheric MTF}
\label{atmomtf}

The variation in phase of a wavefront, $\phi$, between
two points in the pupil plane of a telescope separated by a distance
$\mathbf{r} = \mathbf{r}_1 - \mathbf{r}_2$ can be described by the
structure function
\begin{equation}
\label{strucfunc}
D_\phi (\mathbf{r})= \langle [ \phi(\mathbf{r}_1) - 
\phi(\mathbf{r}_2) ]^2 \rangle,
\end{equation}
where $\langle ... \rangle$
denotes a time average.  
We reserve lower case, e.g., $\phi$, to refer to quantities in the
spatial domain, and upper case, e.g., $\Phi$, for the Fourier
transformed quantity in the frequency domain.  We use $\mathbf{x}$ and
$\mbox{\boldmath$\nu$}$ to refer to spatial coordinates and spatial
frequencies in the image plane and $\mathbf{r}$ and
$\mbox{\boldmath$\kappa$}$ to refer to spatial coordinates and spatial
frequencies in the pupil plane. Thus, the relation between position in
the pupil plane and spatial frequency in the image plane is
$\mathbf{r} = \lambda f \mbox{\boldmath $\nu$}$.

The structure
function can be written in terms of the 
covariance function, which in
turn forms a Fourier pair with the two-dimensional phase-error power
spectrum, $|\Phi(\mbox{\boldmath $\kappa$})|^2$, where
\boldmath $\kappa$ \unboldmath is spatial frequency
in the pupil plane.  
The notation
$|\Phi|^2$
denotes the 
atmospheric phase error power spectrum, which is shorthand
for the expectation value
$\langle \Phi \Phi^* \rangle $, 
where $*$ denotes the complex conjugate.
Consequently, the structure function and the power
spectrum are related by the transform
\begin{equation}
D_{\phi}(\mathbf{r}) = 2\int|\Phi(\mbox{\boldmath$\kappa$})|^2 
     [1-\cos(2\pi\mathbf{\mbox{\boldmath $\kappa$} \cdot r})] d
     \mbox{\boldmath $\kappa$},
\end{equation}
where the integral is taken over the infinite domain.  In the case of
an isotropic power spectrum integration over
the angular part yields
\begin{eqnarray}
\label{DofR}
D_{\phi}(r) & = &  2 \int_{0}^{\infty} 
|\Phi(\kappa)|^2 \;\kappa d \kappa
     \int_{0}^{2\pi} d \theta \; 
     [1-\cos (2\pi \;kr \cos\theta)] \nonumber \\
            & = &  4 \pi \int_{0}^{\infty}
                   |\Phi(\kappa)|^2 
	           [1-J_0(2\pi\;\kappa r)]\kappa \; d \kappa ,
\end{eqnarray}
where $\kappa = |\mbox{\boldmath $\kappa$}|$ \citep{tatarski}, and
$J_0$ is the Bessel function of the first kind.  Note that this
transform, as quoted by \citep[Eq. 4.46]{hardy}, contains a
typographical error.

For Kolmogorov turbulence, with an infinite outer scale and an
inner scale of zero, the two dimensional phase error power spectrum in
the pupil plane is 
\begin{equation}\label{kol_unmod}
|\Phi(\kappa)|^2 = 0.0229\; r_0^{-5/3} \kappa^{-11/3}
\end{equation}
\citep{noll_76}.  The corresponding phase structure function is

\begin{equation}\label{Dtheory}
D_{\phi}(r) = 6.88 \left( \frac{r}{r_0} \right) ^{5/3} ,
\end{equation}
where $r$ is the separation between two points in the pupil
plane and $r_0$ is a scale length known as the Fried parameter
\citep{fried_65}.
The phase structure function
evaluated at $r = \lambda f \nu = D_{tel} \nu_n$
yields the atmospheric MTF
structure function 
\begin{equation}\label{DtoT}
T_{\phi}(\nu_n) = 
\exp \left[
\textstyle 
- \frac{1}{2} D_{\phi}(D_{tel} \nu_n)\right]
\end{equation}
\citep{fried_66}.

\subsection{MTF of an unaberrated pupil}

In the case of a unaberrated, unobscured, circular pupil, the OTF
reduces to the real, circularly symmetric MTF,
\begin{equation}
T_{pup}(\nu_n) = \frac{2}{\pi}
\left[
\cos^{-1} \nu_n - \nu_n(1-\nu_n^2)^{1/2}
\right].
\label{pupmtf}
\end{equation}
In the more general case of a pupil function with central obscuration,
secondary spiders and segment gaps, the MTF can be computed
numerically from the autocorrelation of the pupil function.

\subsection{MTF of an ideal pixel }
\label{pixelmtf}

The MTF of a ideal pixel of dimensions $a \times a$ can be
derived by considering it as a uniformly illuminated
square aperture.  The MTF
is then given by the Fourier transform of this function, which in
normalized frequencies is
\begin{equation}\label{pixelmtfeq}
T_{pix}(\nu_n) = \mbox{sinc}(\pi \nu_n / n_{\rm pix}),
\end{equation}
where $n_{\rm pix}$ is the number of pixels per length $\lambda
F$. For Nyquist sampling, two pixels per $\lambda F$, 
$n_{\rm pix}  =2 $.
For the narrow field camera in Keck/NIRC2 
$n_{\rm pix} = 3.4$ at $H$ band.

\subsection{PSF, encircled energy, and Strehl ratio}
\label{psfee}

The final step involves
finding the 
PSF, $p(r_n)$, and the encircled
energy, $e(r_n)$,
from the Hankel transforms of 
the system MTF $T_{sys}(\nu_n) = T_\phi(\nu_n) T_{pup}(\nu_n)T_{pix}(\nu_n)$.
The Hankel transform of
zero order yields the PSF, 
\begin{equation}
\label{TtoPSF}
p(r_n) = 2\pi \int_0^1 T_{sys}(\nu_n) J_0 ( 2 \pi \nu_n r_n ) 
\nu_n\; d\nu_n 
\end{equation}
where $r_n = r/F\lambda = \theta D_{tel} / \lambda$ is the normalized 
distance in the image plane, 
and the Hankel transform of first order
yields the 
the encircled energy
\begin{equation}
\label{TtoEE}
e(r_n) = \int_0^{r_n} p(r'_n) \; 2 \pi r'_n \;dr'_n = 
2 \pi r_n \int_0^{1}  T_{sys}(\nu_n) J_1(2 \pi r_n \nu_n ) \; d\nu_n.
\end{equation}
where $J_0$ and $J_1$ are the Bessel functions of the first and second
kind, respectively.
The Strehl ratio is defined as $p_{sys}(0)/p_{pup}(0)$, where $p_{pup}$ denotes
the PSF for the unaberrated system. Thus,
using Eq.~(\ref{TtoPSF}) 
\begin{equation}
\label{strehlratio}
S = 
\frac{\displaystyle\int_0^1 T_{sys}(\nu_n)  
\nu_n\; d\nu_n }
{ \displaystyle \int_0^1 T_{pup}(\nu_n)  
\nu_n\; d\nu_n}.
\end{equation}


\clearpage

\end{document}